\documentclass{statsoc}

\usepackage{natbib}

\title[``Additivity'' versus ``Maxitivity'' at the heart of the paradoxical and efficient nature
of Statistics]{``Additivity'' versus 
``Maxitivity'' at the heart of the \linebreak paradoxical and efficient nature
of Statistics} 
\author[M. R\'emon]{M. R\'emon}
\address{Department of Mathematics, Namur University, B-5000 Namur, Belgium 
\\ marcel.remon@fundp.ac.be}

\begin{document}

\begin{abstract}

%
Unlike the Probability Theory based on additivity, Statistical Inference seems to hesitate between 
"Additivity" and a so-called "Maxitivity" approach.
After a brief overview of three types of principles for any (parametric) statistical theory and 
the proof that these principles are mutually exclusive, the paper shows that
two kinds of support measures are conceivable, an {\em additive} one and a {\em maxitive} one (based 
on maximization operators).  Unfortunately, none of them is able to cope with the ignorance part of the 
statistical experiment and, in the meantime, with the partial information given through the structure of the 
data.  To conclude, the author promotes the combined use of both approaches, as an efficient 
middle-of-the-road position for the statistician. 
%
%
%

{\bf R\'esum\'e.} 

Contrairement \`a la th\'eorie de probabilit\'e qui est fond\'ee sur l'additivit\'e, l'inf\'erence 
statistique semble h\'esiter entre "l'Additivit\'e" et ce que d'aucun appelle la "Maxitivit\'e".
Apr\`es un bref survol des trois cat\'egories de principes applicables \`a toute th\'eorie (param\'etrique) 
statistique et la d\'emonstration que ceux-ci sont mutuellement exclusifs, le papier montre que deux types de 
mesure de support sont envisageables, \`a savoir une mesure {\em additive} ou une mesure {\em maxitive} 
(bas\'ee 
sur des op\'erateurs de maximisation).  Malheureusement, aucune n'est capable d'appr\'ehender correctement 
la part d'ignorance contenue dans l'exp\'erience statistique et, dans le m\^eme temps, l'information 
partielle d\'elivr\'ee par la structure des donn\'ees.  En conclusion, l'auteur propose l'utilisation 
combin\'ee des deux approches comme une position efficace et m\'ediane pour le statisticien.
\end{abstract}

\bibliographystyle{Chicago}

\section{Introduction} 

In this paper, we restrict
ourselves to parametric statistical models. Doing that, we know that we leave aside a large part of statistics.  But we think that equivalent reflections can be done for non parametric statistics.
Each statistical theory
tries to answer the same basic question : {\em ``What can we say about the underlying hypotheses, from the  observed data information we get ?''}  The different schools of inference have succeeded
in giving an answer to the question, as long as one accepts some
principles related to these schools. Classical approach is best if one
looks for long-run properties. Bayesian inference should be chosen if
one has meaningful proper prior information over the parameter space.
Structural inference is to be used for transformation models, etc. But
there is not always evident {\em prior} to choose and the Bayesian
approach is therefore difficult to apply; or the data come from a
unique and non replicable experiment and the Classical approach is no longer appropriate. \\[12pt]  
It
would be naive to believe that a single inference theory could be
suitable for all inference problems. In that sense, we totally agree
with \citet{Kalbfleisch70}~: {\em ``In fact, the
main criticism  to be directed at the study of statistical inference
today is the slavish  adherence to rigid dogmas and principles (e.g. Bayes
theorem, likelihood  principle, admissibility, etc.) which is
characteristic of the various schools of inference $\cdots $ To claim that
all problems of inference have been, or even can be, solved by one
overriding principle seems to us naive.''} \\[12pt]  
This
appears to close definitively the search for a general statistical 
inference school. What we try to do in this paper, is to go deeper in 
the formal understanding of such a failure.  We do that by focusing on
the opposition between the way to handle ignorance on one side and
structural data information on the other side.  For that, it is important to look
at the principles underlying the various schools of inference.    
\section{Three sets of principles : a brief overview of Statistical principles}
\subsection{Notations}  
\noindent We represent a
parametric statistical experiment by means of the following model : 
\begin{eqnarray*}  
& - \mbox{} {\cal
M}(X,\Theta,p_{\theta}(x),\mu(x)) \;\; \mbox{where} \; & X \;\; \mbox{is the
sample space,} \\  
& &  \Theta \;\; \mbox{the
parameter space,} \\                         
& &  p_{\theta}(x) \;\; \mbox{the density family with respect to $\mu(x)$, and}
\\ 
& & \mu(x) \;\; \mbox{a $\sigma$-finite measure
over $X$} \\   
& & \mbox{(usually the Lebesgue or the  
countable measure)} \\  
& - \mbox{} \mbox{Plus the
knowledge of the } & \!\!\!\!\!\mbox{observed data ``$x \in E$''.} 
\end{eqnarray*}  

\noindent Many principles will not be
mentioned here because they are mere consequences of general principles
such as the Likelihood or Invariance ones. We think, for instance, of
the Mathematical Equivalence principle (\citet{Birnbaum64}),
which states that our inference  should be independent from any
one-to-one transformation of the sample  space. This principle is a
corollary of the Likelihood principle.  
\subsection{First set : invariance concerning the parameter space}
\subsubsection{The (Strong)
Invariance principle ${\cal I}$}  
\noindent ${\cal I}$ : Let ${\cal M}_{1}(X,\Theta,p_{\theta}(x),\mu(x))$ and    
${\cal M}_{2}(X^{'},\Theta^{'},p^{'}_{\theta^{'}}(x^{'}), 
\mu^{'}(x^{'}))$ be two different models for the same experiment,
connected  by two functions $f : X
\rightarrow X^{'} $ and  $g : \Theta \rightarrow \Theta^{'}$  such that 
$p_{\theta}(\{x:f(x)\!=\!x^{'}_{o}\}) =  p^{'}_{g(\theta)}(x^{'}_{o})$ and
$\mu(\{x:f(x)\!=\!x^{'}_{o}\}) =  \mu^{'}(x^{'}_{o})$ $\forall \theta \in
\Theta$ and $\forall x^{'}_{o} \in X^{'}$. The Invariance principle states
that equivalent inference about $g(\theta)$ should be made from the first
model given the knowledge $``f(x)=x^{'}_{o}"$ as from the second model
given the same observation $``x^{'}=x^{'}_{o}"$. \\[12pt]  
To
better understand the Invariance principle, let us consider 
$N(\mu,\sigma^{2}),$ the Normal
model. Suppose we only observe the value of the standard
deviation $s^{2}=s^{2}_{o}$. The Invariance principle states that
inference about $\sigma$ given the observed $s^{2}_{o}$ should be the
same whether one uses the Normal model $N(\mu,\sigma^{2})$ or the
Chi-square distribution for
$\frac{s^{2}} {\sigma^{2}}$. The Invariance principle requires that
statistical inference does not depend on the choice of parameterization
for the model. A consequence of this principle is the invariance of
inference under one-to-one transformation of the parameter. This
principle is advocated by many authors, see for instance \citet{Hartigan67}. It is at the heart of the paradoxes studied by
\citet{Dawid73} against Bayesian and Structural
inference.  See also  the old Bertrand-Von Mises paradox about the choice of the ratio ``wine-water'' versus ``water-wine'' as our parameterization within an uniform model (\citet{VonMises39}).   
\subsection{Second set : invariance concerning the sample
space}  
\subsubsection{The Censoring principle ${\cal CE}$}  
\noindent ${\cal CE}$ : For any specified outcome $x_{o}$ of an experiment   
${\cal M}(X,\Theta,p_{\theta}(x),\mu(x))$, our statistical evidence is
fully characterized by the function $p_{\theta}(x_{o})$, $\theta\in
\Theta$, without further reference to ${\cal M}$ or $x_{o}$, i.e. all
our statistical information is contained in the likelihood function
(\citet{Birnbaum64}).\\[12pt]
${\cal CE}$ was first proposed by \citet{Pratt62} 
by means of an example : if an accurate voltmeter gave a
reading of 87, does it matter, for the interpretation of this reading
(assumed error-free), whether the meter's range was bounded by 1,000 or
by 100 ? 
\subsubsection{The Stopping Rule principle ${\cal ST}$}  
\noindent ${\cal ST}$ : The Stopping Rule
principle states that the sampling design is irrelevant to statistical
inference at the stage of data analysis. \\[12pt]
This principle is formally
equivalent to the Censoring principle, if one considers the following 
stopping rule : stop the experiment as soon as $``x_{o}"$ is
observed.
${\cal ST}$ can be accepted if one is working
with an experiment which will be performed once only. It is certainly
not a satisfying principle for long-run sampling experiment, which is
the basis of classical inference.  
\subsubsection{The (Strong) Likelihood principle
${\cal L}$}  
\noindent ${\cal L}$ : Suppose a statistical experiment is
characterized by two different models with common parameter space : 
${\cal M}_{1}(X,\Theta,p_{\theta}(x),\mu(x))$ and ${\cal
M}_{2}(X^{'},\Theta,p^{'}_{\theta}(x^{'}),\mu^{'}(x^{'}))$ such that
$p_{\theta}(x_{o}) = c  \cdot p^{'}_{\theta}(x^{'}_{o}) $ for each
$\theta$ in $\Theta$, for some $x_{o}$ in $X$, $x^{'}_{o}$ in $X^{'}$
and for constant
$c \neq 0$. Then ${\cal L}$ states that the same inference should be made
about $\theta$ whatever $x_{o}$ or $x^{'}_{o}$ is observed.\\[12pt]
The Likelihood principle says that all the relevant information for
inference about $\theta$ is contained in the sole knowledge of the
relative likelihood function.
This principle is advocated and criticized
by many statisticians. See for instance \citet{Fisher34},  
\citet{Birnbaum62,Birnbaum64},
\citet{Barnard67,Barnard73}, \citet{Basu73}, \citet{Berger85} or  \citet{Berger88}.
\citet{Birnbaum64} proved that  ${\cal L}$ is equivalent to the
Sufficiency  principle ${\cal S}$ (see next section)  \& ${\cal CE}$ or
to ${\cal S}$ \& ${\cal ST}$. 
\subsection{Third set : Reduction-type principles}   
\subsubsection{The Reduction principle ${\cal R}$}  
\noindent ${\cal R}$ : In logic, if $A \Rightarrow C$
and $B \Rightarrow C$, then $ (A \cup B) \Rightarrow C$. In statistics,
one  has a similar principle. Let $I(A)$ be the inferential information
contained in the observation $A$ [or some statistical inference made from
$A$]. If the data $A$ and  
$B$ lead to the same inference $I(A)=I(B)$, one
should perform equivalent inference from the observation of $A \cup B$ : 
i.e. if $A \Rightarrow I(A)$ and $B \Rightarrow I(B)\!=\!I(A)$,  then $(A
\cup B) \Rightarrow I(A \cup B)\!=\!I(A)\!=\!I(B)$.\\[12pt]
This
Reduction principle was proposed by \citet{Dawid77}. It gives a
general framework for all the (partial) Sufficiency or Conditionality
principles.
\subsubsection{The Sufficiency principle
${\cal S}$}    
\noindent ${\cal S}$ : In an experiment ${\cal M}(X,\Theta,p_{\theta}(x),\mu(x))$, we get the   
same  
information about
$\theta$, if we observe the realization $x_{o}$ or  
only its realization
through a sufficient statistic $T(x_{o}) = t_{o}$.  \\[12pt]
This
principle is certainly the most widely accepted principle   
in statistics. 
Together with ${\cal CE}$ or ${\cal ST}$,  
it implies that   
the likelihood
function is only relevant for inference  
up to a proportional constant.  
\subsubsection{The Conditionality principle ${\cal CO}$}     
\noindent ${\cal
CO}$ : Suppose we have an experiment ${\cal  
M}(X,\Theta,p_{\theta}(x),\mu(x))$  
and a maximal ancillary statistic $T(x)$. 
$T$ is ancillary  
if $p_{\theta}(T(x))$ is independent of $\theta$.  Then our  
inference about $\theta$ should be done through the conditional  
probability
$p_{\theta}(x|T(x))$. \\[12pt]
${\cal CO}$ was studied, among others,
by \citet{Cox58}  
and \citet{Barndorff71,Barndorff73}.  
\citet{Birnbaum62} proved that the Sufficiency 
principle ${\cal S}$ together with the Conditionality principle  
${\cal CO}$
implies the Likelihood principle ${\cal L}$.  
\subsubsection{The
Partial Nonformation principles ${\cal PN}$}  
\noindent ${\cal PS}$
[Partial Sufficiency principles] : Let 
$T(x)$ be a partial sufficient
statistic, in some  
spe\-cified sense, like {\it B-, S-, M-, K-, I-}   
or
{\it L-}sufficiency.  See \citet{Barndorff71}, \citet{Remon84},  
\citet{Cano89} or  \citet{Jorgensen93} for definitions of partial  
sufficiency.  
All the Partial
Sufficiency principles state that   
one gets the same  
inferential information
about some parameter  
of interest from the knowledge of  $``x_{o}"$ or  
$``T(x_{o})"$, and that one has to do inference through  
the marginal
distribution $P_{\theta}(T(x))$.\\[12pt]    
Equivalent Partial
Conditionality principles ${\cal PC}$  
require that our  
inference should be
done through the conditional   
distribution given the observation of some {\it
B-, S-,}  
... ancillary statistic.    
\citet{Barndorff78}
introduced the  
concept of nonformation which generalizes both notions  
of
partial sufficiency and partial ancillarity, and   
leads to Partial
Nonformation principles ${\cal PN}$.  
\subsection{Summary}   
\noindent All these statistical principles can be summarized in  
three principles :   
\begin{itemize}  
\item[$\bullet$] The {\em Invariance principle  
${\cal I}$} about the choice of parameterization;   
\item[$\bullet$] The {\em Likelihood principle
${\cal L}$} about the choice for the
reference sample space, which is equivalent to the   
Censoring
${\cal CE}$   
or Stopping Rule principle ${\cal ST}$  
 together with the Sufficiency
principle ${\cal S}$;  
\item[$\bullet$] The {\em Reduction principle ${\cal R}$} which generalizes the Partial Nonformation principles ${\cal PN}$ about  
the kind of information one has to consider in the data   
($``x_{o}"$ or 
$``T(x_{o})"$ ?).
\end{itemize}   
The next section will discuss the logic of {\em ignorance} versus {\em structural information}  with respect 
to the best choice for a support measure over the hypotheses space $\Theta$.
\section{Is there a good choice for a support measure with respect to the hypotheses ?} 
\subsection{Introduction}    
\noindent In this paper, we choose the general terms  of {\em support measure} to express the support the observation data give to some unknown hypothesis.  When looking for {\em support measures} in the theories of ignorance or uncertainty, one finds a lot of propositions.  Let us just mention here the ancient Laplace's \citeyearpar{Laplace12} inverse probability theory (see \citet{Dale99} or \citet{Fienberg06}), Dempster-Shafer's belief function \citep{Shafer76}, the classical Bayesian {\em a posteriori} probability, the structural inference \citep{Fraser68}, the theory of possibility \citep{Zadeh78,Dubois07}, the plausibility measures \citep{Friedman95} or the recent general uncertainty theory  \citep{Zadeh05}. 
\subsection{The case of the non informative Bayesian priors}    
%
\noindent It is well known that
additive priors, as proposed by the Bayesian theory, are not  
suitable for expressing absence of knowledge about 
hypotheses.  See \citet{Shafer76} : if we have no information about three hypotheses $H_1,H_2$ and $H_3$, we cannot say that we have a better knowledge about $H_1 \cup H_2$ with respect to $H_3$ because we can add these small pieces of (non)information.    
Another example is the one proposed by  
\citet{Bernardo79} :  
we toss a coin and we wish 
to do inference about its bias through the parameter of  
interest $\phi =
|\theta - \frac{1}{2}|$, where $\theta$ is  
the probability of observing {\it
``Head"}.  
We know that the coin is either fair
($H_{1}:\theta\!=\!\frac{1}{2}$),  double-headed ($H_{2}:\theta\!=\!1$) 
or double-tailed ($H_{3}:\theta\!=\!0$).  
We observe $x_0 = ${\it ``Head"}$ \;\cup $
{\it ``Tail"}, i.e. we have no  
information coming from the data.  
The
likelihood function is then  
$l(\theta|x_0) = 1 \hspace{0.3cm} \forall \theta \in
\Theta$.  
If we express our ignorance  
about $\theta$ through an additive uniform
measure~: $p(H_{i})=\frac{1}{3},  
i\!=\!1,2,3$, we therefore state that the
hypothesis $H_{1}:\phi\!=\!0$   
is twice less likely than  
the hypothesis
$H_{A} : \phi\!\neq\!0$.  This contradicts the situation of ignorance.   \\[12pt]
\noindent \citet{Dawid73}  and \citet{Stone76} have proposed many paradoxes against the additive nature of 
the Bayesian prior, especially  in the context of lack of information. \citet{Jeffreys39} worked a lot to 
find non informative Bayesian priors.   In his paper about the history of Bayesian Inference, 
\citet{Fienberg06} writes that trying  to {\em 'derive ``objective'' priors that expressed ignorance or lack 
of knowledge'} is like trying {\em 'to  grasp the holy grail that had eluded statisticians since the days of 
Laplace'}.  In fact, we can broaden the  scope of the incompatibility between ignorance and additivity, to 
situations where partial information is available, i.e. to any kind of support measure.
\subsection{The incompatibility between  ``additivity'' and the logic of ignorance}
\noindent Our knowledge ({\em a priori} or {\em a posteriori}) about the ``true'' unknown hypothesis $\theta_0 \in \Theta$ can be in some way informative.  This does not mean that our  support measure
about this 
hypothesis behaves like a probability measure. Let us define the {\it support
measure} describing the likelihood the  
observation $E$ gives to the hypothesis
$\theta \in \Theta_{1}$ by  $S[\theta \in \Theta_{1}|E]$.  
A support measure, like any plausibility measure \citep{Friedman95}, has to satisfy three ``axioms"
:   
\begin{eqnarray*}  
& \bullet & S[\theta \in \Theta_{1}|E]=0 \; \mbox{if} \; E
\Rightarrow   
(\theta \not\in \Theta_{1})  \\  
& \bullet & S[\theta \in
\Theta_{1}|E]=1 \; \mbox{if} \; E \Rightarrow   
(\theta \in \Theta_{1}) \\  
& \bullet & \Theta_{2} \subseteq \Theta_{1} \; \Rightarrow \;  
S[\theta \in
\Theta_{2}|E] \leq S[\theta \in \Theta_{1}|E] \\
& & \mbox{   
[monotonicity of the support
function]}  
\end{eqnarray*}  
%
%
\noindent The problem for choosing a support measure on $\Theta$ is that this measure should always handle a part of ignorance. Indeed, even when it is an {\em a posteriori} support measure over $\Theta$, there will be  hypotheses $\theta_i$ with equivalent support from the observed data (through the likelihood function, for instance), and the support measure will have to manage this ignorance between these $\theta_i$.  Once again, like in the Bernardo's coin example, this cannot be done by an additive support measure.  Let us prove this incompatibility as a consequence of the Invariance ${\cal I}$ and Likelihood ${\cal L}$ principles.  \\[12pt]
\noindent Suppose that we express our
statistical information   
about $\theta$ by means of an additive posterior support
measure $S[\theta|E]$. Because of the Invariance ${\cal I}$ and Likelihood ${\cal L}$, 
two $\theta$-values, $\theta_{1}$ and $\theta_{2}$,
having the same  
relative  
likelihood cannot be distinguished.  To prove
that,  
one has just to consider the function $g(\theta)$ used  
in ${\cal I}$
as the permutation of $\theta_{1}$   
and $\theta_{2}$.  
${\cal I}$ implies that   
$\mu_{1}\equiv S[\theta_{1}|E] = S[\theta_{2}|E] \equiv \mu_{2}$.  The logic of ignorance  
requires equivalent
inference for $\theta_{1}\cup \; \theta_{2}$  
as for $\theta_{1}$.  
Considering 
$S[\theta|E]$ as additive, one gets  :  
$S[\theta_{1}\cup\theta_{2}|E]=\mu_{1}+\mu_{2}=S[\theta_{1}|E]=\mu_{1}$. 
Hence, $\mu_{1}=\mu_{2}=0$, which is far from convincing.  
All this reasoning
about the consequences of ${\cal I}$  
and ${\cal L}$ was already mentioned, in
similar terms, by  
\citet{Hartigan67}.  \\[12pt]
\noindent As it is clear that additive support measures are incompatible with ${\cal I}$ and ${\cal L}$, one can think that non-additive support measures, like the ones proposed in the possibility theory, will be the correct choice.  The next section shows that support measures built on maximization (or minimization) are also to be questioned. 
\subsection{The case of the possibility measure and its ``Maxitivity'' property }  
%
\noindent The theory of possibility, as well as the theory of plausibility, proposes measures defined in terms of {\em maximization} or {\em minimization}.  \citet{Dubois07} have introduced the pretty terms of {\em ``Maxitivity''} and {\em ``Maxitive measure''} in reference to the additivity property of probability measures.  For instance, the {\em possibility measure} for the state $A \subseteq S$ is denoted by $\Pi(A)$ and defined by :
\begin{eqnarray*}
\Pi(A)& = & \sup_{s \in A} \pi (s) \\
& &  \mbox{where $\; \pi : S \rightarrow [0,1]$ is a possibility distribution for the states $s \in S$} 
\end{eqnarray*} 
\noindent A {\em necessity measure} can be defined for $A \subseteq S$ by $N(A) = \inf_{s \in A} (1 - \pi(s)) = 1 - \Pi(A)$.   One get the following {\em ``maxitivity''} properties :
\begin{eqnarray*}
 \Pi(A \cup B) & = & \max (\Pi(A),\Pi(B)) \\
 N(A \cap B) & = & \min (N(A),N(B)) 
\end{eqnarray*}
\noindent See \citet{Dubois07} and \citet{Sigarreta07} for detailed explanations about possibility and plausibility measures.  Let us note that $\Pi^*(A)\equiv(\Pi(A)+N(A^c))/2$ is still a possibility measure with the additional properties that $\Pi^*(A)=0$ is equivalent to the impossibility of A, and  $\Pi^*(A)=1$ to the certainty of A.  This can be interesting for comparison with {\em a posteriori} Bayesian probability, but it will not be used here.
\subsection{The impossibility of a
 ``maxitive'' support measure satisfying  
$\; {\cal I}, {\cal L}$ and ${\cal R}$} 
%
%
\noindent Let us define our support measure in the framework of the theory of possibility, but in relation to the relative likelihood function $l(\theta;E)$ :
\begin{eqnarray*}
S[ \theta \in \Theta_0|E] & = & \sup_{\theta \in \Theta_0} l(\theta;E) \\
& = & \frac{\displaystyle \sup_{\theta \in \Theta_0} p_{\theta}(E)}{\displaystyle \sup_{\theta \in \Theta} p_{\theta}(E)}
\end{eqnarray*}
It is clear that such a possibility measure satisfies the Invariance and Likelihood principles. However, the Reduction principle is not satisfied.  Indeed, such a support measure based on the sole likelihood function is incompatible with the Reduction principle, as far as partial sufficiency principle is concerned.  Let us consider the Bernardo's \citeyearpar{Bernardo79} coin example again. \\[12pt]
\noindent  Remember that our parameter 
of interest is $\phi = |\theta - \frac{1}{2}|$ where $\theta$ is  
the probability of observing {\it ``Head"}, and that the coin is known to be 
either fair ($H_{1}:\theta\!=\!\frac{1}{2}$),  
double-headed ($H_{2}:\theta\!=\!1$) 
or double-tailed ($H_{3}:\theta\!=\!0$). This time, we 
observe $x_0 = ${\it ``Head"}.  The
likelihood function is   $l(\theta|x_0) = \theta \;\; \forall \theta \in
\Theta$. So, by definition of our support measure, one gets :
\begin{eqnarray*}  
S[\theta|\mbox{{\it ``Head"}}] = 1 - S[\theta|\mbox{\it ``Tail''}]  = & \theta  & \\
 S[H_1 |\mbox{\it ``Head''}]  = S[H_1 |\mbox{\it ``Tail''}] = & \frac{1}{2}  \;\; & \mbox{[The support measure for fairness]}  \\
S[H_2 \cup H_3 |\mbox{\it ``Head''}]  = S[H_2 \cup H_3 |\mbox{\it ``Tail''}]  = & 1  \;\; & \mbox{[The support measure for unfairness]} 
\end{eqnarray*}
\noindent We see that the likelihood, as well as our {\em
  ``maxitive''} support measure, puts its highest support towards the
unfairness of the coin, whatever the first toss gives as a result.  We see also that there is an invariant structure in the model, concerning our parameter of interest.  Indeed, the minimal G-sufficient statistic with respect to $\phi$,  see
\citet{Barnard63}, is  $ T(\mbox{{\it
``Head"}})=T(\mbox{{\it ``Tail"}})=T(\mbox{{\it ``Head'' or  ``Tail"}})$.  Thus, 
from ${\cal R}$ (or ${\cal PS}$) and the fact that the {\em Marginal} likelihood $l(\theta|\mbox{\it ``Head'' or ``Tail''})=1 \;\; \forall \theta \in \Theta$, we should
have  :  
\begin{eqnarray*}  
S[H_{1}|\mbox{{\it ``Head"}}]=S[H_{1}|\mbox{{\it ``Tail"}}] \stackrel{\cal R}{=} S[H_{1}|\mbox{{\it ``Head" or
``Tail"}}] = 1 \\
S[H_{2}\cup H_{3}|\mbox{{\it ``Head"}}] =
S[H_{2}\cup H_{3}|\mbox{{\it ``Tail"}}] \stackrel{\cal R}{=} 
S[H_{2}\cup H_{3}|\mbox{{\it ``Head" or ``Tail"}}] = 1   
\end{eqnarray*}
\noindent This is clearly a better situation in terms of inferential support, as the first toss of a coin gives no information at all about the fairness or unfairness of a coin.  \\[12pt]
This example shows the impossibility for a {\em ``maxitive''} support
measure to satisfy the Reduction principle, as this last one, through
structural invariance and partial sufficiency, introduces  {\em
  Marginal likelihood function} in the scene of inference.  And
therefore, an {\em additive} operation in terms of
likelihood. Moreover, as we observed in the coin example, {\em single} and {\em marginal} likelihood functions can express totally different support with respect to the hypotheses. Which one should we prefer ?  Our choice will, {\em de facto}, contradict either the Reduction or the Likelihood principle.
\subsection{Summary}
%
%
\noindent  In this section, we have seen that neither the {\em
  additivity} nor the {\em maxitivity} approach is ``the'' solution
for our support measure.  The first one cannot handle properly the
ignorance present in any statistical problem, while the second one
cannot cope with its structural invariance (for instance, a location-scale structure emerging with the asymptotic normal model when the number of observations increases). 
\begin{itemize}
\item[$\bullet$] The {\em additive} 
Bayesian posterior approach satisfies the Likelihood ${\cal L}$ and Reduction  
${\cal R}$ principles, but not the Invariance ${\cal I}$ principle. ${\cal R}$ will be valid under the condition that a 
reference  
parameterization is chosen as well as a proper prior distribution over the parameter space $\Theta$.   
This {\em extra information} is required if paradoxes are to be avoided (\citet{Dawid73,Stone76}).
\item[$\bullet$] The {\em maxitive} 
Maximized or Profile Likelihood approach (\citet{Barndorff94})  satisfies the Invariance ${\cal I}$ and Likelihood 
${\cal L}$ principles, but not the Reduction ${\cal R}$ principle.  The
Generalized Likelihood Ratio tests are based on this type of support  
measure, as well as the Maximum Likelihood point estimation.
The {\em extra
information} needed  
here to avoid paradoxes is the
long-run
behavior of the model, as well as its structural invariance (\citet{Stein56, Barnard65, Berger88}).
\item[$\bullet$] The {\em mixed additive-maxitive} Marginal or Conditional Inference approaches have not yet been considered in this paper. The Marginal (or Conditional)
Likelihood approach is defined in  
the same way as the Profile Likelihood
approach,  
but using the marginal [respectively conditional]  
likelihood function 
$l(\theta;T(x))$ [$l(\theta;x|T(x))$] instead of the simple likelihood function
$l(\theta;x)$, as we did in Bernardo's example.    The Invariance ${\cal I}$ and Reduction ${\cal R}$ principles will be satisfied here, but not the Likelihood ${\cal L}$ principle.
The problem here is the definition  
of what is a partial
nonformative statistic $T(x)$ (\citet{Barndorff78,Remon84,Cano89,Zhu94,Barndorff94}). 
The use of a marginal or conditional likelihood function requires {\em extra information} as the
knowledge of the stopping rule  
because the  
marginal or conditional density can differ from
one stopping rule to another.  
${\cal L}$ is no
longer valid for this type of inference.
\end{itemize}
\noindent Bayesian, Profile Likelihood and Marginal/Conditional Likelihood
inferences are 
three  
major approaches corresponding to the  possible 
{\em two-by-two} combinations of our general principles.  Other  
inference  
methods can be classified in the same way,
depending on the list of  
principles they  
satisfy. But none will be able to satisfy all these principles, as there is an internal incompatibility between them. This incompatibility can be seen as a dilemma between an {\em additive} or a {\em maxitive} approach for dealing with the ignorance and the structural information contained in the data.    
\section{Conclusions : the dilemma between {\em ``Additivity''} and {\em ``Maxitivity''}}  
\noindent Our point of view is that discussion about Statistical Schools of Inference should not focus so much on the kind of principle one keeps or rejects, or even  by-passes thanks to some well chosen {\em  extra information}. Indeed, any inference theory seems to miss some
information, as  extra information is always needed to avoid paradoxes. 
Statisticians should be more aware of and worried by the mathematical properties of the support measure they wish to use.  Here comes the dilemma between the {\em ``additivity''} and the {\em ``maxitivity''} of our support measure. \\[12pt]
\noindent One can think that most statisticians will prefer an
additive approach, by similarity with the probability theory, but this
is not so clear.  Indeed, the core of the point estimation is done in
a {\em maxitive} environment. And if they have to compare hypotheses, they will normally use likelihood ratios, which are based on {\em maxitive} support measures.  We think that neither  the {\em maxitive} nor the {\em additive}  approach should be promoted as the sole possible approach.  \\[12pt]
\noindent Our point of view is that statisticians should use both perspectives, in a dialogal process, like 
in the Marginal/Conditional Likelihood approach.  EM algorithm (\citet{Dempster77}) is also a good example of 
this combined use of {\em ``maxitive''} and {\em ``additive''} operators.  This double nature of the support 
measure is, for us, the {\em characteristic of  statistical inference}, as statisticians should consider 
themselves as staying in the middle of the road,  trying to reconcile the logic of ignorance (related to {\em 
``Maxitivity''}) and the logic of information (linked to {\em ``Additivity''}). This is also the source of 
the efficiency of many statistical {\em ad hoc} methods.     %
\bibliography{mybib} 

\end{document}